\documentclass{ws-p10x7}
\begin{document}
\title{
\vskip-2cm
\rightline{\normalsize FERMILAB--Conf--00/252--E}
\rightline{\normalsize CMU--HEP--00--04}
\rightline{ }
Recent results from Selex
}
\author{
\small
The SELEX Collaboration \\ 
J.~Russ$^{3}$,
G.~Alkhazov$^{11}$,
A.G.~Atamantchouk$^{11}$,
M.Y.~Balatz$^{8}$$^{,\ast}$,
N.F.~Bondar$^{11}$,
P.S.~Cooper$^{5}$,
L.J.~Dauwe$^{17}$,
G.V.~Davidenko$^{8}$,
U.~Dersch$^{9}$$^{,\dag}$,
A.G.~Dolgolenko$^{8}$
% ...
}
\address{
% ...
$^3$Carnegie-Mellon University, Pittsburgh, PA 15213, U.S.A.\\
% ...
$^5$Fermilab, Batavia, IL 60510, U.S.A.\\
% ...
$^8$Institute of Theoretical and Experimental Physics, Moscow, Russia\\
$^9$Max-Planck-Institut f\"ur Kernphysik, 69117 Heidelberg, Germany\\
% ...
$^{11}$Petersburg Nuclear Physics Institute, St. Petersburg, Russia\\
% ...
$^{17}$University of Michigan-Flint, Flint, MI 48502, U.S.A.\\
% ...
}

\maketitle
\footnotetext[1]{Talk given by J.~Russ at the 
XXXth International Conference on High Energy Physics
July 27 - August 2, 2000, Osaka, Japan. Proceedings to be published by
World Scientific.}

\twocolumn[
\abstract{
\vskip-2cm
The SELEX experiment (E781)  is 3-stage magnetic spectrometer for
the study of charm hadroproduction at
large $x_{F}$ using 600 Gev $\Sigma ^{-}$, $\pi ^{-}$ and $p$ beams. 
New precise measurements of the $\Lambda _{c}$, $D ^{0}$, and $D _{s}$ 
   lifetimes are presented.
   We also report results on $\Lambda _{c}$ and $D_s$ production 
   by $\Sigma ^{-}$, $\pi ^{-}$ and $p$ beams at $x_{F}>0.2$.  The
   data agree with expectations from color-drag models to explain charm
   particle/antiparticle production asymmetries. 
}]
\footnotetext{$^{\ast}$deceased}
\footnotetext{$^{\dag}$Present address: SAP, Walldorf, Germany}
\footnotetext{$^\ddag$Now at Imperial College, London SW7 2BZ, U.K.}
\footnotetext{$^\S$Now at Instituto de F\'{\i}sica da Universidade Estadual de Campinas, UNICAMP, SP, Brazil}
\footnotetext{$^\P$Now at Physik-Department, Technische Universit\"at M\"unchen, 85748 Garching, Germany}
\footnotetext{$^\parallel$Present address: The Boston Consulting Group, M\"unchen, Germany}
\footnotetext{$^{\ast\ast}$Present address: Lucent Technologies, Naperville, IL}

\section{Introduction}

Charm physics explores QCD phenomenology in both perturbative 
and nonperturbative regimes. Charm lifetime measurements test models 
based on $1/ M_{Q}$ QCD expansions and evaluate corrections from 
non-spectator W-annihilation and Pauli interference to perturbative QCD 
matrix elements.  Production studies test leading order (LO)
and next to leading order (NLO) perturbative QCD.
The parton-level processes are symmetric between c and $\overline{c}$,
but fixed-target data on hadrons show significant asymmetries.
Experimental data from different incident hadrons ($\pi$, $p$ and
$\Sigma ^{-}$) may help to illuminate hadron-scale physics.~\cite{Frix97}
  
\section{Features of the Selex spectrometer}

The SELEX experiment at Fermilab is a 3-stage magnetic spectrometer
~\cite{spec}.  The negative Fermilab Hyperon Beam at 600 GeV
had about equal fluxes of $\pi^-$ and $\Sigma^-$.  The positive beam was
92\% protons. For 
charm momenta in a range of 100-500 Gev/c mass resolution is constant
and primary (secondary) vertex resolution is typically 270 (560) $\mu$m.
A RICH detector labelled all particles above 25 GeV/c, greatly reducing 
background in charm analyses.~\cite{RICH}
Our charm analysis requires: (i) primary/secondary vertex separation  
$L \ge 8 \sigma $ ($\sigma$ is the combined error); (ii) the 
decay tracks extrapolated to the primary 
vertex z position to miss such that the second-largest transverse miss
distance $\ge$ 20$\mu$m; (iii) the secondary
vertex to lie outside any charm target by at least 0.5 mm; 
(iv) decays to occur within a given fiducial region; and
(v) proton and kaon tracks to be identified by the RICH.

\footnotetext{$^{\dag\dag}$Now at Instituto de F\'{\i}sica Te\'orica da Universidade Estadual Paulista, S\~ao Paulo, Brazil}
\footnotetext{$^{\ddag\ddag}$Present address: SPSS Inc., Chicago, IL}
\footnotetext{$^{\S\S}$Now at University of Alabama at Birmingham, Birmingham, AL 35294}
\footnotetext{$^{\P\P}$Present address: DOE, Germantown, MD}
\footnotetext{$^{\ast\ast\ast}$Present address: Siemens Medizintechnik, Erlangen, Germany}
\footnotetext{$^{\dag\dag\dag}$Present address: Deutsche Bank AG, Eschborn, Germany}
\section{Measurements of the $\Lambda _{c}$, $D ^{0}$ and $D _{s}$ lifetimes}
The lifetime hierarchy for
the charm system presents a challenge to HQET and pQCD methodologies due
to the low charm quark mass.  We report here
lifetimes for the following charm hadrons in the
decay modes listed: (i) $\Lambda _{c} ^{+}  \rightarrow p K ^{-} \pi ^{+}$;
(ii) $D ^{0}  \rightarrow  K ^{-} \pi ^{+}$ and 
$K ^{-}  \pi ^{+} \pi ^{-}\pi ^{+} $ + c.c.; and
(iii) $D _{s}  \rightarrow  K ^{*}(892) K$ and  $\phi \pi $.

SELEX charm signals are extracted by the sideband subtraction method
with a fixed signal region ($\pm 2.5\sigma_M$, i.e., 20 MeV/$c^2$) 
centered on the 
charm mass.  Sidebands of equal width are defined above and below
the charm mass region.  The background under the charm peak is the average
of the two sideband regions.

$\pi /K$ misidentification causes mixing of charm signals.  In SELEX
this is significant only for the $D_{s}$ peak.  
For both $D_s$ modes kaon momenta are $\le$160 Gev/c to reduce 
misidentification.  
Any KK$\pi$ event having a pseudo-$D^{\pm}$ mass in the
interval 1867 $\pm$ 20 Mev/$\rm{c}^{2}$ is removed to eliminate an 
artificial lengthening of the $D_s$ lifetime, even though some 
of these are real $D_s$ events.

We make a binned maximum 
likelihood fit simultaneously to signal
and sideband regions in reduced proper time  
$ t^{*} ={M(L-8 \sigma)/ p c }$.  SELEX results are given in 
Table~\ref{tab:life}.  The $D_s$/$D^0$ lifetime
ratio R, sensitive to the W-annihilation amplitude in the
weak decay,  is 1.17 $\pm$ 0.057, consistent with other
recent published results.

\begin{table}
\caption{PRELIMINARY SELEX lifetimes (fs)} \label{tab:life}
\begin{tabular}{|c|c|c|c|} \hline
Charm Particle & $\tau$ & $\sigma_{stat}$ & $\sigma_{syst}$\\ \hline 
$\Lambda _{c} ^{+}$ & 198.1 & 7.0 & 5.5  \\
 $D ^{0}$ &  407.0 & 6.0 & 4.3  \\
 $D_{s}$  & 475.6 & 17.5 & 4.4  \\ \hline
\end{tabular}
\end{table}

\section{Charm and Anticharm Hadroproduction}

\begin{figure}
\epsfxsize120pt
\figurebox{}{}{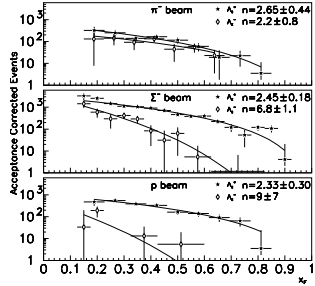}
\caption{$x_F$ dependence of $\Lambda_c^+$ and $\overline{\Lambda_c}^-$
production by different beams} \label{fig:lcxf}
\end{figure}

For $\Sigma ^{-}$ and proton beams (valence q, not $\overline{\rm{q}}$) 
${\Lambda _{c}}^{+}$ production
is favored over ${\overline{\Lambda} _{c}}^{-}$at all $x_F$, and the
difference increases dramatically at large $x_F$.   The efficiency
difference for charm and anticharm hadrons in SELEX is small, at most 3\%.   
$\pi^-$ production (valence q and $\overline{\rm{q}}$) shows a small asymmetry 
that becomes consistent
with zero at large $x_F$.  The data are shown in Fig.~\ref{fig:lcxf}. 
For $D_s$ the $\Sigma^-$ beam produces far more $D_s^-$ (s-quark) than 
$D_s^+$ ($\overline{s}$ quark), with the difference
increasing at large $x_F$.  For pions, the yield is smaller and
the integral yield difference between $D_s^-$ and $D_s^+$ for $x_F \ge$ 0.2
is consistent with zero.  These asymmetry patterns suggest a connection 
between a charm or anticharm quark and remnants of the
beam fragmentation jet, like in the quark-gluon string model.
There is no evidence in the large-$x_F$ events for a diffractive
partner to the leading charm hadron, again consistent with a color-drag
picture.

The $p_T$ distributions for $\Lambda_c^+$ production from the three different
beam hadrons have identical gaussian distributions out to $p_T^2 \le$ 2
$\rm{(GeV/c)}^2$, after which the good-statistics $\Sigma^-$ data show a
change to power-law behavior, as is expected from QCD and has been 
observed in D-meson production by $\pi^-$. 

\section{Summary}
SELEX has explored charm hadroproduction in the large $x _{F}$ 
region using different beams.  Results favor the color-drag explanation
of production asymmetry.  We used the data to measure preliminary charm 
lifetimes: 
$\tau(\Lambda _{c})=198.1 \pm 7.0 \pm 5.5$ fs,
 $\tau(D ^{0})=407.0 \pm 6.0 \pm 4.3$ fs, and
$\tau(D_s)$ of $475.6 \pm 17.5 \pm 4.4$ fs.  
The SELEX result for the ratio
$\tau _{D_{s}}/{\tau_{D^{0}}}$ is 1.17 $\pm$ 0.057, 3.3$\sigma$ from unity.

\end{document}